\documentclass[prl,twocolumn,superscriptaddress,showpacs,%
preprintnumbers,amsmath,amssymb,floatfix]{revtex4}
\usepackage{times}
\usepackage{graphicx}
\usepackage{dcolumn}
\usepackage{bm}
\usepackage{subfigure}

\newcommand{\tr}{\mathrm{Tr}}
\newcommand{\vect}[1]{\bm{#1}}
\begin{document}

\title{Nonperturbative Treatment of Double Compton 
Backscattering in Intense Laser Fields}
\author{Erik L{\"o}tstedt}
\email{Erik.Loetstedt@mpi-hd.mpg.de}
\affiliation{Max-Planck-Institut f\"{u}r Kernphysik, 
Postfach 103980, 69029 Heidelberg, Germany}
\author{Ulrich D.~Jentschura}
\affiliation{Department of Physics, 
Missouri University of Science and Technology, 
Rolla, Missouri 65409-0640, USA}
\affiliation{Institut f\"ur Theoretische Physik, Universit\"at Heidelberg, 
Philosophenweg 16, 69120 Heidelberg, Germany}

{\begin{abstract}
The emission of a pair of entangled photons by an electron in an intense laser
field can be described by two-photon transitions of laser-dressed, relativistic
Dirac--Volkov states.  In the limit of a small laser field intensity, the
two-photon transition amplitude approaches the result predicted by double
Compton scattering theory.  Multi-exchange processes with the laser field,
including a large number of exchanged laser photons, cannot be described
without the fully relativistic Dirac--Volkov propagator. The nonperturbative
treatment significantly alters theoretical predictions for 
future experiments of this kind. We quantify the degree of polarization
correlation of the photons in the final state by employing the well-established
concurrence as a measure of the entanglement. 
\end{abstract}}

\maketitle

\begin{figure}[t]
\includegraphics[width=1\columnwidth]{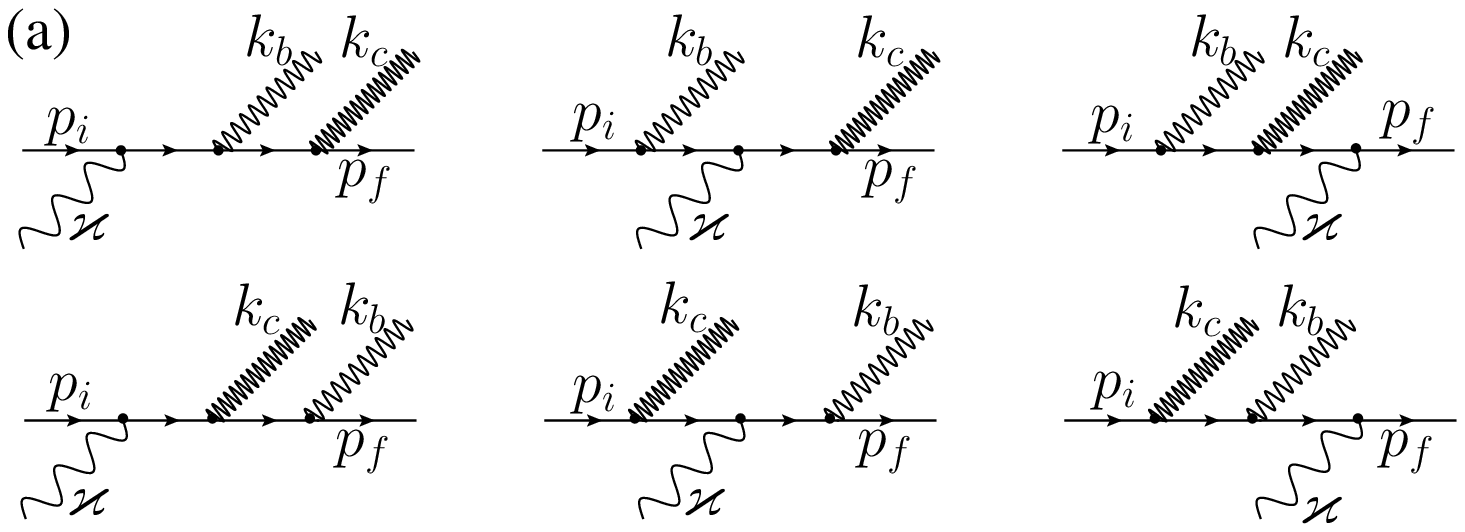}
$\quad$\\
\includegraphics[width=1\columnwidth]{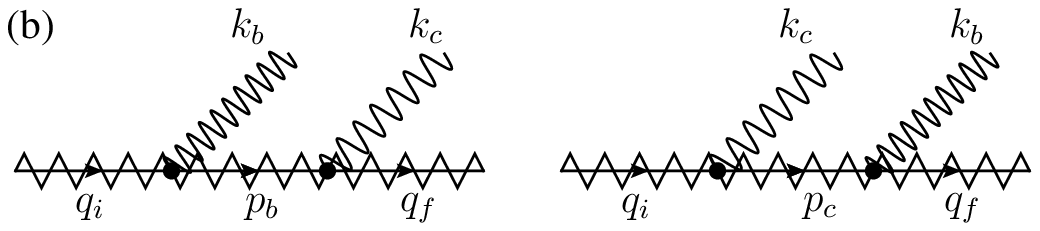}
\caption{The Feynman diagrams for perturbative (a), and 
nonperturbative (b) double Compton scattering. In (a),
an electron with initial momentum $p_i$ absorbs one laser photon
$\varkappa$, emits two photons with
wave vectors $k_b$ and $k_c$ and ends with final momentum $p_f$.
Instead in (b), the laser-dressed initial state with average four-momentum $q_i$, decays to the
final state with average four-momentum $q_f$ under emission of two
photons.  In the intermediate state the momenta are labeled
by $p_b$ and $p_c$, respectively.  The nonperturbative interaction with the
laser field is depicted by dressing the electron lines with a zigzag line.
\label{Feynman_graph}} 
\end{figure}

{\em Introduction.---}In ordinary  Compton scattering~\cite{KleinNishina1929},
a photon is scattered inelastically by an electron. For
photons with energy much less than the electron's rest mass, the quantum
mechanical expression for the cross section agrees with the one obtained by
classical electrodynamics. Nonlinear Compton scattering is
encountered when several photons from a strong laser beam are scattered by a
free electron to produce a photon of different energy; this process has been
calculated theoretically~\cite{BrKi1964,NaFo1996} and successfully
measured~\cite{BaBoetal2004,BabIl2006}.  Recently, there has been an increased
interest in a different nonlinear generalization of Compton scattering where a
free electron collides with a laser pulse and emits {\it two} photons at the
same time. This process has no classical counterpart, and indeed, as we will
see, the two photons exhibit a paradigmatic quantum feature: namely, their
polarizations are entangled. Properly optimized, two-photon emission from
backscattering of laser photons at an electron beam holds the promise of
providing entangled light at much larger energy than conventionally used for
quantum information purposes~\cite{Zeil1999}.

With relativistically strong lasers being available in many laboratories
worldwide, the current record being a laser intensity of $10^{22}$~W/cm$^2$ at
the focus~\cite{Yanovskyetal2008}, the quest for observing genuine
laser-induced quantum effects in the relativistic regime continues. However,
the peak field strengths are still orders of magnitudes below the quantum electrodynamic
(QED) critical
field $E_c=-m^2/e=10^{16}$~V/cm for pair creation (here $m$ and $e=-|e|$ denote
the mass and charge of the electron, respectively, and we use natural
relativistic units $c=\hbar=\epsilon_0=1$).  Two-photon emission by a
laser-dressed electron via nonperturbative double Compton backscattering is a
strong-field, relativistic quantum effect which could be observed
without the additional complications connected with the ultra-relativistic particle
beams necessary for laser-dressed pair creation~\cite{Re1962,Buetal1997,Mu2009}.

The theory of perturbative double Compton scattering, the reaction in which one
photon scatters on an electron to produce two final photons was calculated by
Mandl and Skyrme~\cite{MandlSkyrme1952}, recently reexamined
in~\cite{Bell2008}, and experimentally confirmed
in~\cite{Cavanagh1952,SaSiSa2008}. The relevant Feynman diagrams
are displayed in Fig.~\ref{Feynman_graph}(a).
 In~\cite{ScScHa2006,ScScHa2008,ScMai2009},
the simultaneous emission of two photons is interpreted in terms of the Unruh
effect.  Other two-photon processes that have been investigated, both
theoretically and experimentally are double
bremsstrahlung~\cite{Baier1981293,KahLiuQuar1992,Korol2006}, two-photon
synchrotron emission~\cite{SoVoletal1976,FoKho2003}, and the total rate of
two-photon emission in a crossed field~\cite{MoRi1974}.  However, the
generalization to nonperturbative double Compton scattering has not been
recorded in the literature to the best of our knowledge.
 
The purpose of this Letter is twofold:  To show {\em (i)} that photon pairs 
with quantifiable entanglement can be produced from double Compton 
scattering in intense fields, and that {\em (ii)} nonperturbative effects have to be
incorporated to make reliable predictions for a relativistically strong laser pulse.

{\em Nonperturbative QED formulation.---}The 
interaction of an electron with a laser field of arbitrary intensity can be
treated in the formalism of strong-field QED (Furry picture), where the
classical external field is included in the unperturbed Hamiltonian. For the
present problem, the calculation of the amplitude of the process amounts to
evaluating the Feynman diagrams shown in 
Fig.~\ref{Feynman_graph}(b). 
The external electron lines are Volkov states $\Psi$, exact solutions to
the Dirac equation with an external laser field,
$(i\hat{\partial}-m-e\hat{A})\Psi=0$, and the propagator is the Dirac-Volkov
propagator \cite{ReEb1966,LoJeKe2008_2}. We denote four-vector scalar products of four-vectors $q$ and $p$
as $q\cdot p=q^\mu p_\mu=q^0p^0-\vect{q}\cdot\vect{p}$, and the
Dirac contraction is written as $\hat{p}=\gamma\cdot p$. 
The laser four-vector potential $A^\mu=a^\mu \cos(\varkappa\cdot x)$ propagates in the negative
$x^3$-direction with wave four-vector $\varkappa$ and frequency $\omega$, and is linearly
polarized in the $x^1$-direction. We also introduce the intensity parameter
$\xi=|e|\sqrt{|a^2|/2}/m$, which can be used to classify the regime of
relativistic laser-matter interaction: $\xi\ll 1$ corresponds to the
perturbative regime, and $\xi\ge 1$ to the nonperturbative. To describe the
emitted photons, we employ spherical coordinates so that the momenta read
$k_{b,c}=\omega_{b,c}(1,\sin\theta_{b,c}\cos\psi_{b,c},
\sin\theta_{b,c}\sin\psi_{b,c},\cos \theta_{b,c})$, with $\omega_{b,c}$ being
the frequency. To study the polarization correlation, we also need a basis for
the polarization vectors of the photon pair: we use
$\epsilon_{b,c}^{\lambda_{b,c}=1}=(0,\cos\theta_{b,c}\cos\psi_{b,c}, \cos\theta_{b,c}
\sin\psi_{b,c},-\sin\theta_{b,c})$, and
$\epsilon_{b,c}^{\lambda_{b,c}=2}=(0,-\sin\psi_{b,c},\cos\psi_{b,c},0)$, so that a
generic polarization vector of the photons is given by
$\epsilon_{b,c}=(c_1\epsilon_{b,c}^1+c_2\epsilon^2_{b,c})/
\sqrt{|c_1|^2+|c_2|^2}$, for some complex constants $c_{1,2}$.  The initial
electron is assumed to propagate in the $x^3$-direction with four-momentum
$p_i=(E_i,\vect{p}_i)$, colliding head-on with the laser pulse. The average, or
quasi momentum \cite{NiRi1964} of the electron immersed in the laser wave is
given by $q_i=p_i-\varkappa e^2a^2/(4\varkappa\cdot p_i)=(Q_i,\vect{q}_i)$, with average
mass $m_\ast=\sqrt{q_i^2}$, and the corresponding quantities for the final
electron are labeled by $p_f$, $q_f$.

Having fixed the notation, we proceed to calculate the 
scattering amplitude
$S$. The calculation follows the usual steps of laser-dressed 
QED~\cite{LoJeKe2008_2}, 
and we present only the final result,
\begin{align}\label{twophoton_S}
& S = i 
\sum_{n=1}^\infty
\sum_{s=-\infty}^\infty
\frac{(2\pi)^4e^2 m \,
\delta^4(q_i-q_f+n\varkappa-k_b-k_c)}{2 V^2\, \sqrt{\omega_c\omega_b Q_i Q_f}} 
\nonumber
\\[2ex]
& \times u^\dagger_{f}\gamma^0 \,
\left[M_{bfc}^{n-s} \frac{\hat{f}_b+m}
{p_b^2-m_\ast^2}M_{ibb}^{s} +{M}^{n-s}_{cfb}
 \frac{\hat{f}_c+m}{p^2_{c}-m_\ast^2}M_{icc}^s\right] u_{i}.
\end{align}
Here $V$ is the quantization volume, 
%
%
$M^N_{jkl}=
A_{0,N}^{jk} 
\hat{\epsilon}_{l}
+A_{1,N}^{jk} 
\left(\hat{\epsilon}_{l}
\frac{e\hat{\varkappa}\hat{a}}{2\varkappa\cdot
 p_{j}}+\frac{e\hat{a}\hat{\varkappa}}{2\varkappa\cdot p_k}
\hat{\epsilon}_{l}\right) -
A_{2,N}^{jk}\,\frac{e^2a^2\varkappa\cdot\epsilon_{l}\hat{\varkappa}}
{2\varkappa\cdot p_j\varkappa 
\cdot p_{k}}$, 
$N$ is an integer, $j,k\in\{i,f,b,c\}$, $l\in\{b,c\}$, 
$p_{b,c}=q_i-k_{b,c}+s\varkappa$, 
$\hat{f}_{b,c}=\hat{p}_{b,c}+
\frac{e^2a^2}{4\varkappa \cdot p_{b,c}}\hat{\varkappa}$, and
$A_{h,N}^{jk}$ is a generalized Bessel function \cite{KoKlWi2006},
%
$A_{h,N}^{jk}=
\int_{0}^{2\pi}
\frac{\cos^h\!\theta}{2\pi}
 e^{iN\theta-i(\alpha_j-\alpha_k)\sin\theta
+i(\beta_j-\beta_k)\sin2\theta}d\theta
$
,
%
$h\in\{0,1,2\}$, 
with the arguments $\alpha_j=ea\cdot p_j/(\varkappa\cdot p_j)$, 
$\beta_j=e^2a^2/(8\varkappa\cdot p_j)$, $j\in\{i,f,b,c\}$.
The spinors $u_{i,f}$ are
normalized according to $u^\dagger_{i,f}\gamma^0u_{i,f}=1$.

The energy-momentum conserving 
delta function contains the integer $n$, which is  the
net number of photons absorbed during the entire collision. The 
second index of summation $s$, which appears in the propagator momenta
$p_{b,c}$, is the net number of photons 
exchanged before emitting the second photon [$k_b$ or $k_c$, 
depending on the diagram, see Fig.~\ref{Feynman_graph}(b)]. 
The amplitude \eqref{twophoton_S} is 
gauge invariant under 
$\epsilon_{b,c}\to\epsilon_{b,c}+\Lambda k_{b,c}$, where $\Lambda$ is 
an arbitrary constant.
Another important aspect of the amplitude $S$ is the possibility 
for the propagator momenta to reach the laser-dressed 
mass shell $p_{c,b}^2=m_\ast^2$, 
which indicates the split up of the process into two sequential
single Compton scattering events \cite{MoRi1974}. At such a resonance,
 where the matrix element formally diverges, 
$S$ may be rendered finite by including a small, 
imaginary  correction to the laser-dressed electron mass and 
energy~\cite{BeMi1976,UDJ2009}, or alternatively 
be regularized with an external parameter 
such as the laser pulse length or a finite detector resolution. In the 
following, we will always consider parameter regions such that the
sequential Compton scattering cascade is forbidden by energy-momentum
conservation or is exponentially suppressed by a large-order Bessel function. 
This selection is in accordance with
planned experiments recently discussed in
Refs.~\cite{BroMarBinColEva2008,Thirolfetal2009}.
In order to facilitate the detection of the rather weak two-photon signal, the 
measurement should be done in energy and angular regions where the 
the single Compton scattering process is strongly suppressed. 

In the following we evaluate the differential rate
\begin{equation}\label{rate}
d\dot{W}=\frac{1}{T}|S|^2 \frac{Vd^3 q_f}{(2\pi)^3}
\frac{ Vd^3 k_b}{(2\pi)^3}\frac{ Vd^3 k_c}{(2\pi)^3}\,,
\end{equation}
where $T$ is the long observation time. Integrating 
over the final electron momentum $\vect{q}_f$ and the final photon
energy $\omega_c$, we end up with the rate 
$d\dot{W}/d\omega_b d\Omega_b d\Omega_c$, differential 
in the directions $d\Omega_{b,c}=d\cos\theta_{b,c}d\psi_{b,c}$ of the two photons and in 
$d\omega_b$. The photon energy $\omega_c$ is given by
\begin{equation}\label{omega_c}\begin{split}
\omega_c &=
\frac{n\varkappa\cdot q_i -k_b\cdot q_i-n\varkappa\cdot k_b}
{n\varkappa\cdot k_c/\omega_c+q_i\cdot k_c/\omega_c
+ k_b\cdot k_c/\omega_c}\\
&\approx \frac{4n\omega E_i-\omega_b
\left[\theta_b^2
E_i+\frac{m^2}{E_i}(1+\xi^2)\right]}{\theta_c^2
E_i+\frac{m^2}{E_i}(1+\xi^2)},
\end{split}
\end{equation}
where $k_c/\omega_c$ is independent of $\omega_c$. The last line 
in Eq.~\eqref{omega_c} holds if 
$n\frac{\omega}{m}\ll\frac{m}{E_i}
\approx \theta_b\approx\theta_c\ll1$, which is the parameter
regime on which we will concentrate (backscattering geometry with 
relativistic electron energy).
Moreover, Eq.~\eqref{omega_c} implies that
the sum of the two photon energies is
limited by $\omega_b+\omega_c\le 4n\omega(E_i/m)^2/(1+\xi^2)$. 

\begin{figure}[tb]
\includegraphics*[width=0.49\columnwidth]{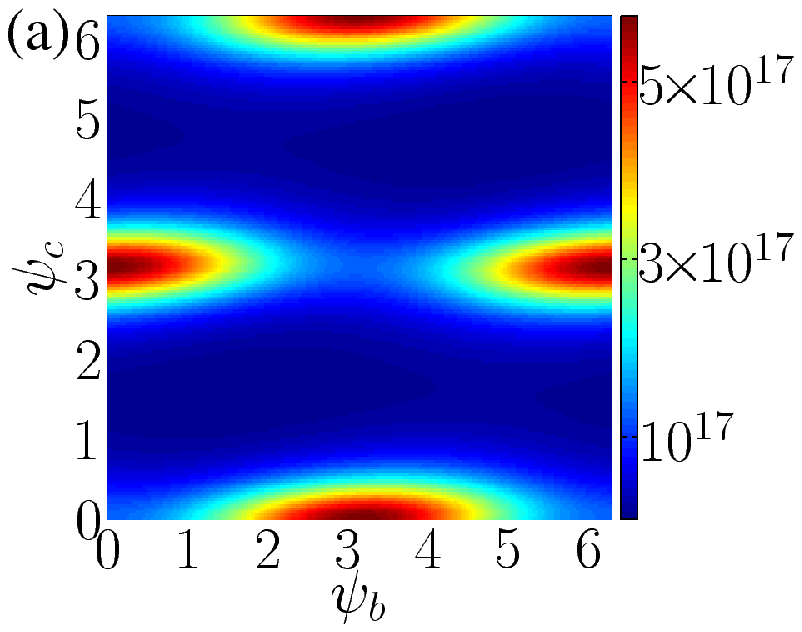}
\includegraphics*[width=0.49\columnwidth]{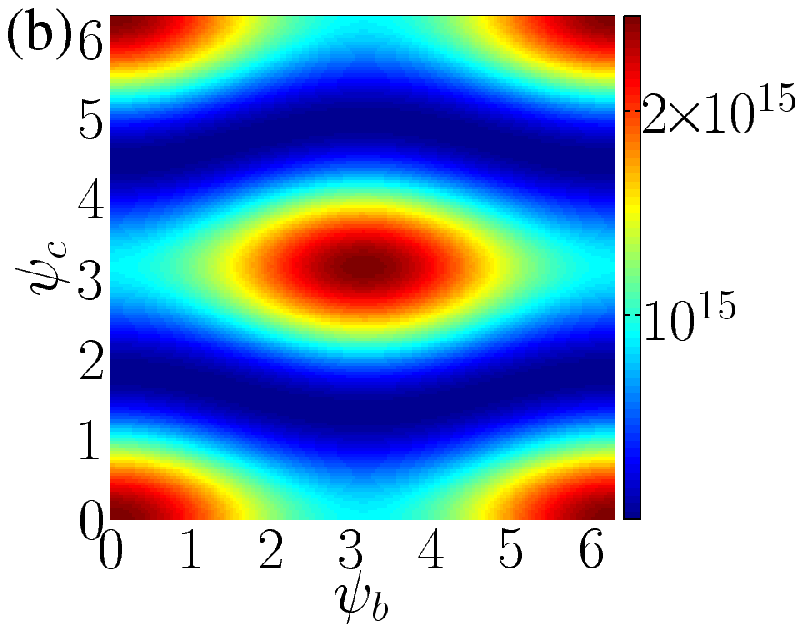}
$\qquad$\\
\includegraphics*[width=0.49\columnwidth]{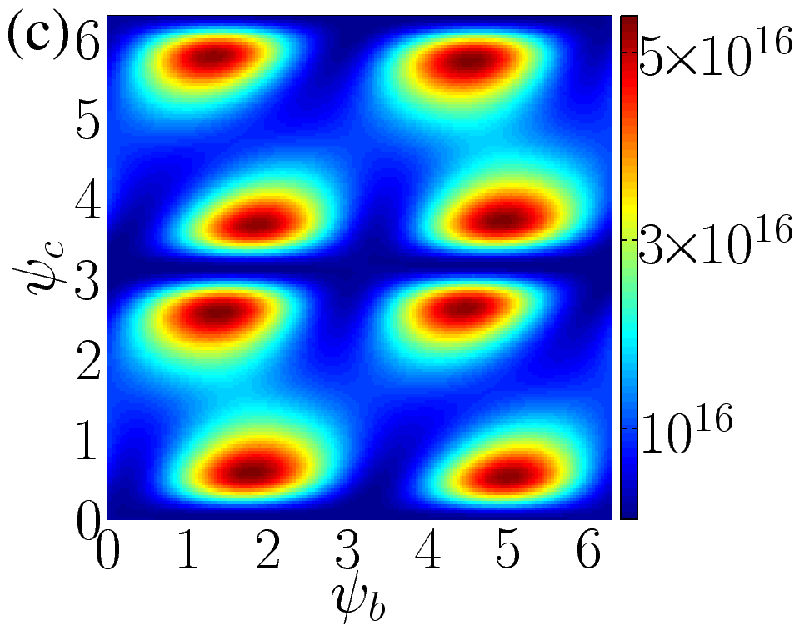}
\includegraphics*[width=0.49\columnwidth]{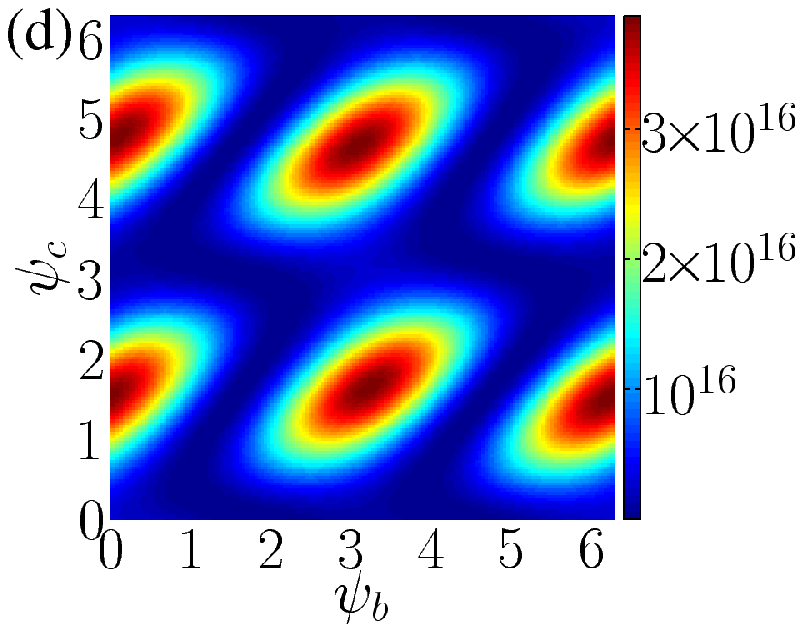}
\caption{
(color).
Comparison of the perturbative and nonperturbative
approach.  Shown is the laboratory frame rate $d\dot{W}/d\omega_b d\Omega_b
d\Omega_c$, in units of s$^{-1}$sr$^{-2}$MeV$^{-1}$, for the nonperturbative
[(a), (c)] and perturbative [(b), (d)] case. The parameters employed are
$E_i=10^3m$, $\omega=2.5$ eV, $\xi=1$, $\omega_b=1$ MeV, and
$\theta_b=\theta_c=10^{-3}$. The photon polarizations are given by
$\epsilon_{b,c}=\epsilon_{b,c}^1$ in (a) and (b), and
$\epsilon_{b,c}=\epsilon_{b,c}^2$ in (c) and (d).  
\label{psibpsicgraph}}
\end{figure}

{\em Calculated differential rate.---}In 
Fig.~\ref{psibpsicgraph}, we show the differential rate in the 
laboratory frame, for a specific set of parameters, and compare
to the corresponding rate obtained from the perturbative 
formula~\cite{MandlSkyrme1952,Bell2008}
which includes only one interaction with the laser field 
[see Fig.~\ref{Feynman_graph}(a)]. We have checked that
the expression~\eqref{rate} agrees with the one obtained in 
\cite{MandlSkyrme1952} in the limit of small laser intensities. 
Since we take the 
initial electron to be relativistic, $E_i=10^3 \, m$, it will emit
mainly in the forward direction, $\theta_{b,c}\sim m/E_i$. The laser
parameters $\omega=2.5$ eV and $\xi=1$ corresponds to an optical laser
with intensity $5.5\times 10^{18}$ W/cm$^2$. Since the quantum 
parameter $\chi=\xi p_i\cdot \varkappa/m^2$~\cite{NiRi1964} is small 
($=10^{-2}$) here, 
spin effects are 
marginal and we therefore average (sum) over the initial (final) 
spin of the electron. The small value of  $\chi$ also permits us 
to neglect effects arising from electron-positron pair creation, 
since the $e^+$-$e^-$ production rates are exponentially 
suppressed. For the parameters used in Fig.~\ref{psibpsicgraph}, 
up to $n=20$ laser photons participate, so that $\omega_c\le60$ MeV
according to  Eq.~\eqref{omega_c}.
The results from 
Fig.~\ref{psibpsicgraph} suggest that the differential rate
varies strongly as a function of the angles and polarization. It 
becomes clear that to interpret data from planned experiments of 
this kind \cite{Thirolfetal2009}, the nonperturbative formula 
\eqref{rate} has to be used. 

\begin{figure}[t]
\includegraphics*[width=1\columnwidth]{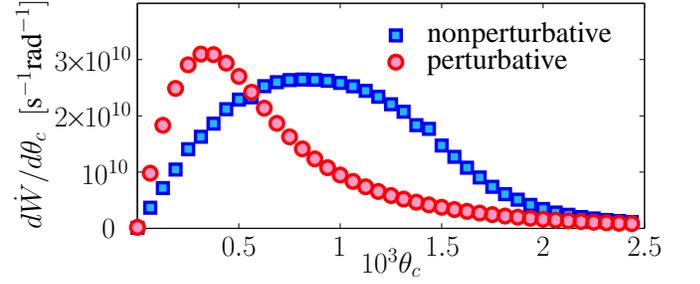}
\caption{(color online). The integrated and polarization-summed 
lab-frame rate $d\dot{W}/d\theta_c=\sin\theta_c d\dot{W}/d\cos\theta_c$, 
differential only in the angle $\theta_c$. Shown are 
results for the perturbative and the nonperturbative calculations, 
for $E_i=10^3m$, $\omega=2.5$ eV, $\xi=1$.
\label{integrated_Rate}}
\end{figure}

To answer the question whether the {\it integrated} nonperturbative rate differ
significantly from the one predicted by the usual double Compton scattering
formula, we show in Fig.~\ref{integrated_Rate} the differential rate
$d\dot{W}/d\theta_c$, integrated over the azimuth angles $\psi_{b,c}$, the
polar angle $\theta_b$ of one of the final photons and the energy $\omega_b$,
and summed over the final photon polarizations.  The energy integration was
limited to the interval between 1 keV and 1 MeV to avoid the infrared
divergence at $\omega_b\to0$~\cite{BrFey1952} and  cascade contributions for 
larger
$\omega_b$, and for the same reason the integration over $\theta_b$ was
performed over the interval $(0,1.5\times10^{-3})$~radians. Restricting the
final phase space in this way ensures that contributions from single Compton
scattering are negligible; at polar angles smaller than $1.5\times10^{-3}$
radians all harmonics occur at energies larger than 1 MeV. Integrating the
nonperturbative curve in Fig.~\ref{integrated_Rate}, we obtain a total rate in
the laboratory frame of $\dot{W}=3.5\times 10^{7}$ s$^{-1}$.  For the
perturbative curve, we get $\dot{W}_{\textrm{pert}}=2.5\times 10^{7}$ s$^{-1}$,
from which we gather that even for the integrated rate, the nonperturbative
corrections are significant. The obtained two-photon rate 
should be compared to the total rate of nonlinear 
single Compton scattering \cite{NiRi1964}, which amounts to 
$3\times10^{13}$ s$^{-1}$ for the same parameters as in 
Fig.~\ref{integrated_Rate}. Employing an electron beam with $10^9$ electrons
per bunch, a laser pulse of duration $100$ fs, photon 
energy $\omega=2.5$ eV,
intensity $5.5\times 10^{18}$ W/cm$^2$ (corresponding to $\xi=1$)
\cite{VULCAN}, and assuming perfect transverse overlap of the two pulses, we
estimate that about $2\times 10^3$ photon pairs per shot may be expected.

\begin{figure}[t]
\includegraphics*[width=0.49\columnwidth]{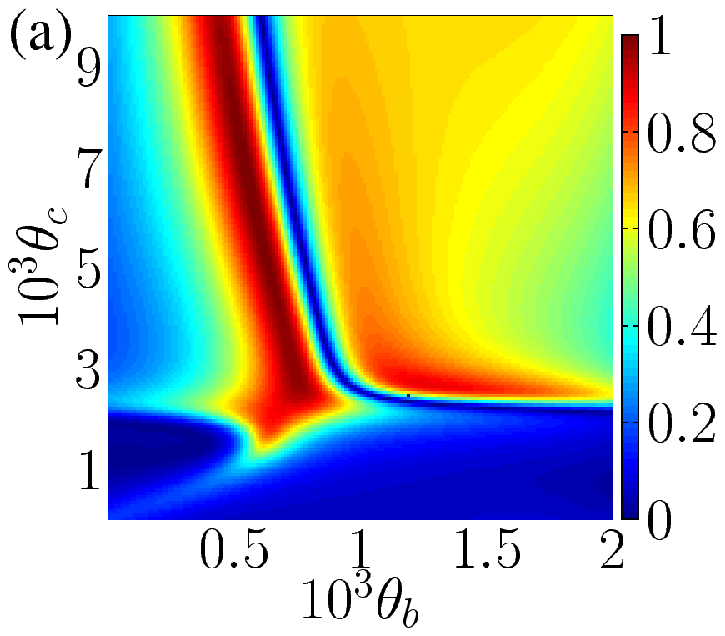}
\includegraphics*[width=0.49\columnwidth]{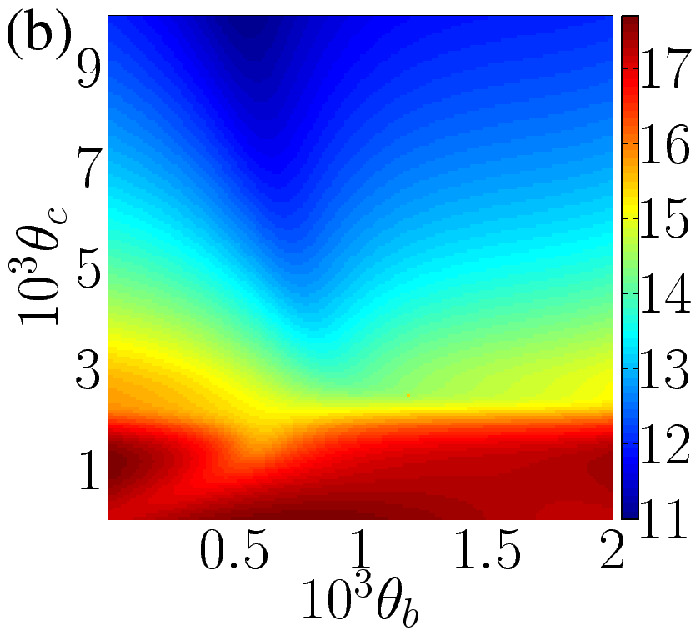}
$\qquad$\\
\includegraphics*[width=0.49\columnwidth]{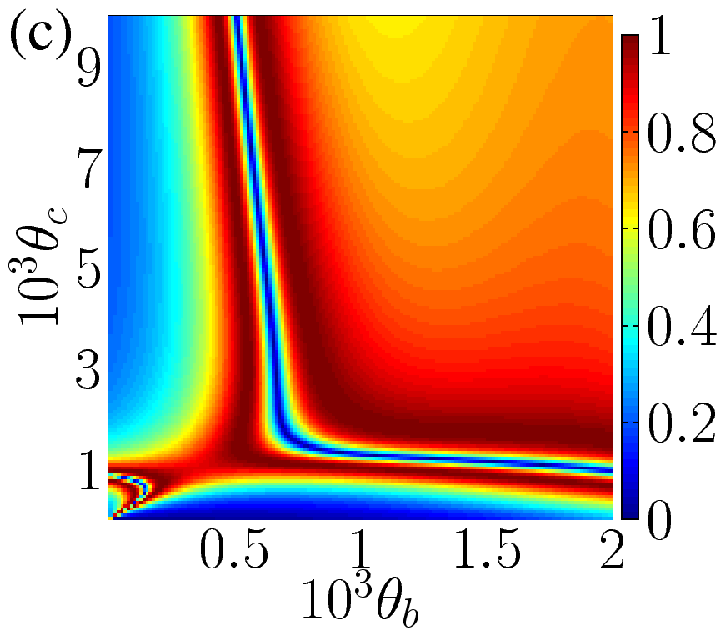}
\includegraphics*[width=0.49\columnwidth]{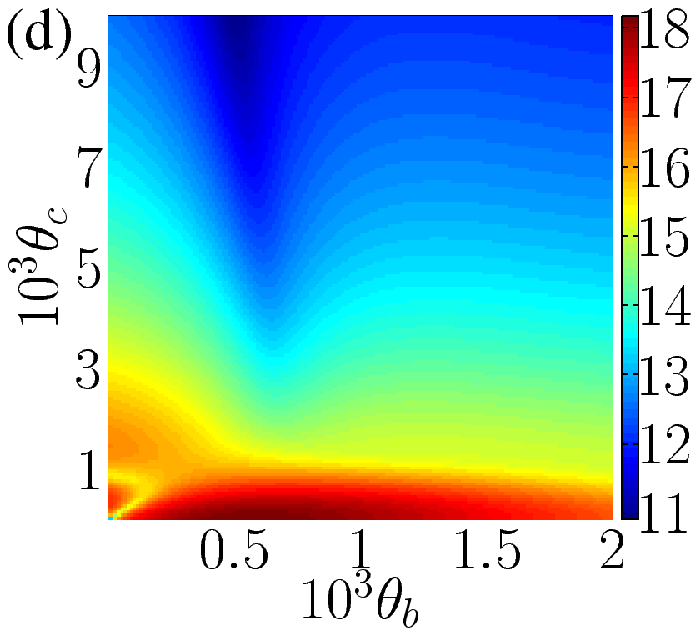}
\caption{
(color).
In panel (a) we show the concurrence $C(\rho_f)$ [see
Eq.~\eqref{Concurrence}], using the nonperturbative expression for the matrix
element, which should be compared to panel (c), where the concurrence is shown
for the perturbative case.  The parameters employed are $E_i=10^3m$,
$\omega=2.5$ eV, $\xi=1$, $\omega_b=1$ MeV, and $\psi_b=\psi_c=0$.  As a
reference, we also display in the right column the logarithm of the
polarization-summed 
differential rate $\log_{10}(d\dot{W}/d\omega_b d\Omega_b d\Omega_c)$, in units
of s$^{-1}$sr$^{-2}$MeV$^{-1}$, for the nonperturbative [(b)] and perturbative
[(d)] case.  \label{diffConcurrence}}
\end{figure}

{\em Entanglement.---}Having investigated 
the differential and total photon pair production rate, 
we now turn to the interesting question of the quantum mechanical 
correlation between the final state photons. To quantify the degree of
polarization entanglement, we employ the well-established 
concurrence $C(\rho_f)$ \cite{Wootters1998} as an entanglement measure.
Assuming an unpolarized initial electron and unobserved final spin, 
we trace out the spin polarizations of the initial and final electron and
calculate the $4 \times 4$ final density matrix $\rho_f$ of the polarizations
$\lambda_{b,c}  \in \{ 1, 2 \}$ of the two emitted photons. Then,
$C(\rho_f)$ is given by
\begin{equation}\label{Concurrence}
C(\rho_f)=\textrm{max}(0,\sqrt{\zeta_1}-\sqrt{\zeta_2}
-\sqrt{\zeta_3}-\sqrt{\zeta_4}),
\end{equation}
where the $\zeta_j$'s are the eigenvalues, in descending order,
of the matrix 
$Q=\rho_f (\sigma_y \otimes \sigma_y) \rho_f^\ast 
(\sigma_y \otimes \sigma_y)$,
where $\sigma_y$ is the second Pauli matrix. For a maximally entangled 
state, $C(\rho_f)=1$, 
and for a non-entangled state $C(\rho_f)=0$. We note that the 
concurrence has recently been used to study correlation in the 
two-photon decay of a bound state \cite{radtke2008}.
For the present case, the 
final density matrix $\rho_f$ can be computed from the normalized 
matrix element \eqref{twophoton_S}. Writing 
$S=S_{r_i,r_f}(\lambda_b,\lambda_c)$,
where $r_i$ ($r_f$) denotes the 
spin polarization of the initial (final) electron, we have for the matrix elements
of $\rho_f$,
\begin{equation}
\langle \lambda_b,\lambda_c |\rho_f |\lambda'_b,\lambda'_c\rangle=
\frac{N}{2}\sum_{r_i,r_f} 
S_{r_i,r_f}(\lambda_b,\lambda_c)
S^\ast_{r_i,r_f}( \lambda'_b,\lambda'_c).
\end{equation}
Here $N$ is a normalization constant, which can be found by requiring
$\tr\,\rho_f=1$.  The concurrence, as defined in Eq.~\eqref{Concurrence}, is a
gauge invariant quantity, furthermore it does not depend on the basis used to
describe the polarization of the photons $k_{b,c}$. $C(\rho_f)$ depends
sensitively on the energy and the directions of the emitted photons. One
example of the fully differential concurrence is displayed in
Fig.~\ref{diffConcurrence}, which  shows the necessity of the nonperturbative
formalism to predict the degree of entanglement.

{\em Conclusions.---}We have studied the process of nonperturbative two-photon decay
of a laser-dressed electron.
Our results significantly alter the theoretical predictions
as compared to a perturbative treatment of the laser;
they lead to novel features in the angular and integral characteristics,
which could be resolved using presently available intense laser 
facilities.

\begin{acknowledgments}
The authors acknowledge support by the National Science 
Foundation and by the Missouri Research Board.
The work of E.L.~has been supported by Missouri University of
Science and Technology.
\end{acknowledgments}

\end{document}